\documentclass[11pt]{article}

\usepackage[final]{acl}

\usepackage{times}
\usepackage{latexsym}

\usepackage[T1]{fontenc}

\usepackage[utf8]{inputenc}

\usepackage{microtype}

\usepackage{inconsolata}

\usepackage{graphicx}

\usepackage{amsmath,amssymb,mathtools}
\usepackage{booktabs}
\usepackage{multirow}
\usepackage{float}
\usepackage{placeins}
\usepackage{algorithm}
\usepackage{algorithmic}
\usepackage{makecell}
\usepackage{comment}

\newcommand{\tok}[1]{\texttt{\char`\<#1\char`\>}}
\newcommand{\spktag}[1]{\texttt{\char`\<#1\char`\>}}

\providecommand{\keywords}[1]{}

\usepackage{xcolor}

\title{G-STAR: End-to-End Global Speaker-Tracking Attributed Recognition}

\author{
Jing Peng$^{2,4,*}$ \quad
Ziyi Chen$^{1,3,*}$ \quad
Haoyu Li$^{1,2,*}$ \quad
Yucheng Wang$^{5}$ \\
Duo Ma$^{4}$ \quad
Mengtian Li$^{3}$ \quad
Yunfan Du$^{3}$ \quad
Dezhu Xu$^{3}$ \quad
Yu Xi$^{2}$ \quad
Kai Yu$^{2}$ \quad
Shuai Wang$^{1,4,\dagger}$ \\
$^1$ Nanjing University \quad
$^2$ Shanghai Jiao Tong University \quad
$^3$ Central Media Technology Institute, Huawei \\
$^4$ Shenzhen Research Institute of Big Data \quad
$^5$ ETH Zürich \\
\texttt{jing.peng@sjtu.edu.cn, zc612997@gmail.com, haoyu.li.cs@sjtu.edu.cn} \\
\small $^*$ Equal contribution. \quad $^\dagger$ Corresponding author.
}


\begin{document}
\maketitle

\begin{abstract}
We study timestamped speaker-attributed automatic speech recognition (SA-ASR) for long-form, multi-party speech with overlap. In this setting, chunk-wise inference must preserve meeting-level speaker identity consistency while producing time-stamped, speaker-labeled transcripts. Prior Speech-LLM systems tend to prioritize either local diarization or global labeling, lacking the ability to jointly model fine-grained temporal boundaries and robust cross-chunk identity linking. We propose \textbf{G-STAR}, an end-to-end framework that couples a cache-conditioned speaker-tracking module with a Speech-LLM transcription backbone. The tracker provides structured speaker cues with temporal grounding, and the LLM generates attributed text conditioned on these cues. G-STAR supports component-wise optimization and joint end-to-end training, enabling flexible learning under heterogeneous supervision and domain shift. Under chunk-wise decoding protocols, experiments on both oracle-segmented local evaluation and full-meeting global evaluation show strong speaker-attributed transcription performance. 
\end{abstract}


\section{Introduction}
\label{sec:intro}

Large language models (LLMs) have recently become central to speech understanding systems, largely owing to their strong semantic reasoning, discourse-level context modeling, and instruction-following capabilities~\citep{tang2023salmonn,chu2023qwen,xu2025qwen3,goel2025audio}. A prevailing paradigm, often referred to as \emph{Speech-LLM}, combines a pre-trained speech encoder with an LLM through a lightweight projector that maps acoustic representations into the LLM embedding space~\citep{peng2024survey_speechllm,cui2025recent,xu2025fireredasr}. This encoder--projector--LLM recipe has shown promising results across a broad range of speech tasks, especially those requiring higher-level linguistic and pragmatic understanding.

However, real-world conversational audio commonly departs from the single-speaker assumption. In cocktail-party conditions and multi-party meetings, overlapping speech and rapid turn-taking demand not only accurate transcription~\citep{cetin06_interspeech,yoshioka18_interspeech, shriberg2001observations}, but also reliable attribution of \emph{who spoke what}~\citep{kanda22b_interspeech}, and increasingly \emph{when}~\citep{shafey19_interspeech}. This motivates \emph{speaker-attributed automatic speech recognition} (SA-ASR), which aims to produce a structured transcript containing both lexical content and speaker attribution~\citep{lu2021streaming,raj2023surt}. In this work, we consider a timestamped SA-ASR formulation: given an input waveform $x$, the system outputs a sequence of attributed segments
\begin{equation}
\label{eq:intro_saasr}
\mathcal{Y}=\{(s_n,\tau^{\mathrm{st}}_n,\tau^{\mathrm{ed}}_n,\mathbf{y}_n)\}_{n=1}^{N},
\end{equation}
where $s_n$ denotes the speaker identity, $(\tau^{\mathrm{st}}_n,\tau^{\mathrm{ed}}_n)$ denotes temporal boundaries, and $\mathbf{y}_n$ is the word sequence. Crucially, for long-form recordings processed in chunks, SA-ASR requires \emph{meeting-level} (or \emph{global}) speaker identity consistency: the same real-world speaker should be assigned the same identity across the entire recording~\citep{raj24_odyssey}, rather than being re-indexed independently in each chunk.

The central challenge is therefore not simply to recognize multiple speakers within a short segment. A practical meeting system must keep speaker identities stable across chunks, retain timestamp-level grounding, and avoid relying on post-hoc global clustering that is difficult to integrate with LLM-style generation. This motivates a tracking-conditioned generation view: speaker tracking should provide structured, reusable cues to the transcription model rather than remain a separate downstream diarization step.

Recent efforts begin to address parts of this problem, yet leave key gaps. SpeakerLM~\citep{speakerlm} augments the Speech LLM pipeline with an additional speaker embedding extractor, enabling robust diarization within a local chunk; however, it does not provide an explicit mechanism to \emph{globally associate} speaker identities across multiple chunks when long recordings must be processed segment-by-segment. JEDIS-LLM~\citep{shi2025jedisllm} introduces a \emph{Speaker Prompt Cache} to store and index speaker representations, thereby enabling the model to emit \emph{globally consistent} speaker labels together with the corresponding transcripts under chunk-wise inference; nevertheless, it lacks explicit perception and emission of fine-grained temporal boundaries (timestamps). TagSpeech~\citep{huo2026tagspeech} further strengthens temporal grounding by incorporating a dedicated speaker encoder and temporal anchors to obtain speaker and timestamp cues, but it similarly does not resolve the \emph{meeting-level} global speaker identity linking problem under chunk-wise inference. Taken together, these advances highlight the need for an end-to-end framework that simultaneously achieves (i) timestamped, overlap-aware attribution, and (ii) persistent, meeting-level global speaker identity consistency.

To bridge these gaps, we propose \textbf{G-STAR: End-to-End Global Speaker-Tracking Attributed Recognition}, an LLM-based end-to-end SA-ASR framework designed for long-form, multi-speaker speech with overlapping regions. G-STAR integrates an interpretable \emph{speaker-tracking module} based on Sortformer~\citep{park2024sortformer,medennikov2025streamingsortformer}-style modeling with a Speech-LLM transcription backbone. The speaker-tracking module provides structured, time-aware speaker cues, while the LLM generates attributed text conditioned on these cues. Unlike cascaded SA-ASR pipelines that separately perform ASR, diarization, timestamp alignment, and post-hoc fusion, G-STAR injects projected, slot-aligned speaker cues into LLM decoding so lexical tokens, timestamps, and global speaker labels are produced under a shared conditioning context. This modular design enables both \emph{component-wise} optimization (improving tracking and transcription separately) and \emph{joint} end-to-end optimization (integrating speaker and lexical modeling in a unified objective), thereby supporting flexible training strategies under dataset imbalance and domain shifts. Most importantly, G-STAR is built to preserve \emph{global} speaker identity consistency across chunk-wise long-form inference, enabling meeting-level speaker labeling without post-hoc global clustering. Accordingly, our contributions are threefold:

\begin{itemize}
  \item We present \textsc{G-STAR}, an end-to-end, LLM-based SA-ASR system for multi-party meetings that produces \emph{timestamped} speaker-attributed transcripts with \emph{meeting-level global speaker identity consistency} under chunk-wise long-form inference. We will release the model and code.
  \item We evaluate \textsc{G-STAR} in both local and global SA-ASR settings on challenging meeting benchmarks, comparing against representative Speech-LLM systems and strong conventional pipelines under clearly specified VAD and collar protocols.
  \item We conduct analyses and ablations to identify the key design factors behind the gains, including speaker cue fusion strategies, fusion ratio choices for injecting speaker cues into LLM generation, and the impact of hierarchical cross-entropy objectives.
\end{itemize}

\section{Related Work}
\label{sec:related}

\begingroup
\setlength{\emergencystretch}{1em}

\subsection{Sortformer and AOSC for Speaker Tracking}

Sortformer~\citep{park2024sortformer} proposes a \emph{tracking-centric} formulation that connects frame-level speaker activity with token-level speaker attribution.
Using notation aligned with Section~\ref{sec:method}, an audio encoder maps each chunk $x^{(t)}$ to acoustic representations $\mathbf{H}^{(t)}=f_{\mathrm{enc}}(x^{(t)})$, and a Sortformer-style tracker estimates speaker activity over arrival-order slots.
Slot $k$ denotes the $k$-th speaker by first appearance, which makes speaker indices interpretable and reduces the label-permutation ambiguity that arises when multi-speaker outputs are estimated independently.
This arrival-order view is particularly useful for chunk-wise processing, where speakers must not be re-indexed independently in each chunk.

To carry this ordering across chunks, Streaming Sortformer~\citep{medennikov2025streamingsortformer} introduces the \emph{Arrival-Order Speaker Cache (AOSC)}.
For chunk $t$, the tracker conditions on the previous cache state and produces frame-synchronous speaker cues,
\begin{equation}
\mathbf{S}^{(t)} = f_{\mathrm{trk}}(x^{(t)}, \mathcal{C}^{(t-1)}),
\end{equation}
after which the cache is updated as
\begin{equation}
\mathcal{C}^{(t)} = \operatorname{UpdateCache}(\mathcal{C}^{(t-1)}, \mathbf{S}^{(t)}).
\end{equation}
Conceptually, AOSC stores compact evidence for speakers in first-arrival order: new speakers receive new slots, while returning or temporarily silent speakers can be linked back to existing slots.
This mechanism (Fig.~\ref{fig:streaming_sortformer}) provides the cross-chunk identity state needed by G-STAR, where speaker tracking serves as structured conditioning for Speech-LLM generation.

\begin{figure}[t]
  \centering
  \includegraphics[width=\columnwidth]{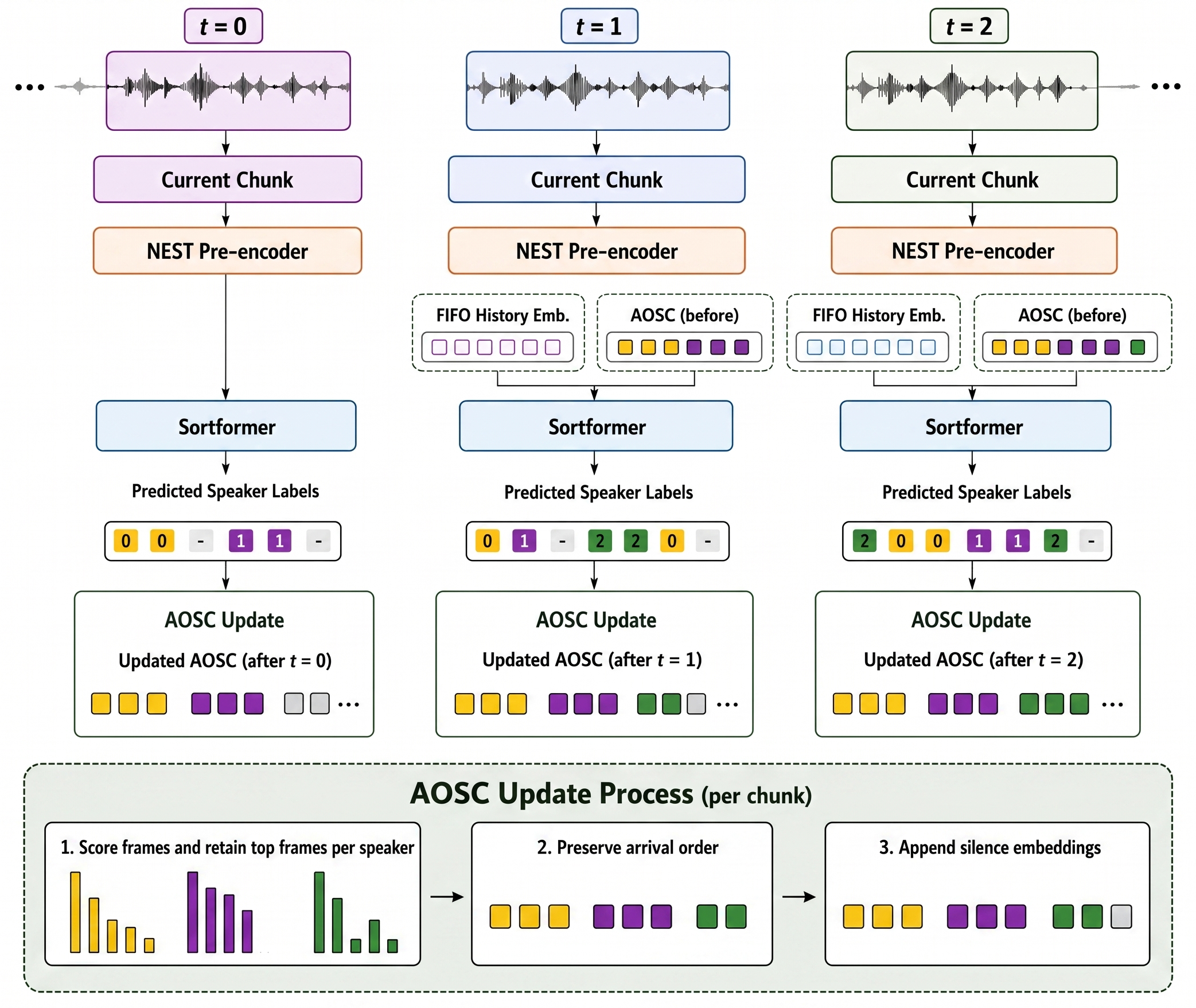}
  \caption{Illustration of Streaming Sortformer with AOSC. The cache stores arrival-order speaker states across chunks and reuses them to keep speaker identities consistent when speakers reappear.}
  \label{fig:streaming_sortformer}
\end{figure}

\subsection{Serialized Output Training} 
Serialized Output Training (SOT)~\citep{kanda2020serialized} addresses the \emph{label permutation} issue in overlapped and multi-speaker transcription by converting multi-stream supervision signals into a single serialized target sequence.
Instead of predicting $K$ parallel speaker-dependent streams, SOT defines a deterministic serialization function that maps the set of speaker-labeled utterances (or token sequences) into a single sequence by inserting explicit \emph{speaker tags} (e.g., \spktag{spk1}, \spktag{spk2}) between lexical tokens~\citep{kanda2020serialized}.
This enables standard autoregressive or sequence-to-sequence training while avoiding explicit permutation search at inference time.

Crucially, SOT provides an \emph{output protocol} for speaker-attributed text that is compatible with LLM-style generation: the model emits a single token stream that interleaves speaker identifiers and words, and can be naturally extended with additional structural tokens (e.g., timestamps) when temporal grounding is required.
However, under chunk-wise inference, SOT alone does not guarantee \emph{meeting-level} identity consistency: speaker tags can still permute across chunks unless the model maintains a persistent speaker concept (e.g., via an explicit tracking state).
Therefore, in our framework, SOT is adopted as the structured output interface, while global identity persistence is provided by cache-conditioned Sortformer-style tracker outputs used as speaker cues.

\begin{figure*}[t] 
  \centering
  \includegraphics[width=\textwidth,height=0.34\textheight,keepaspectratio]{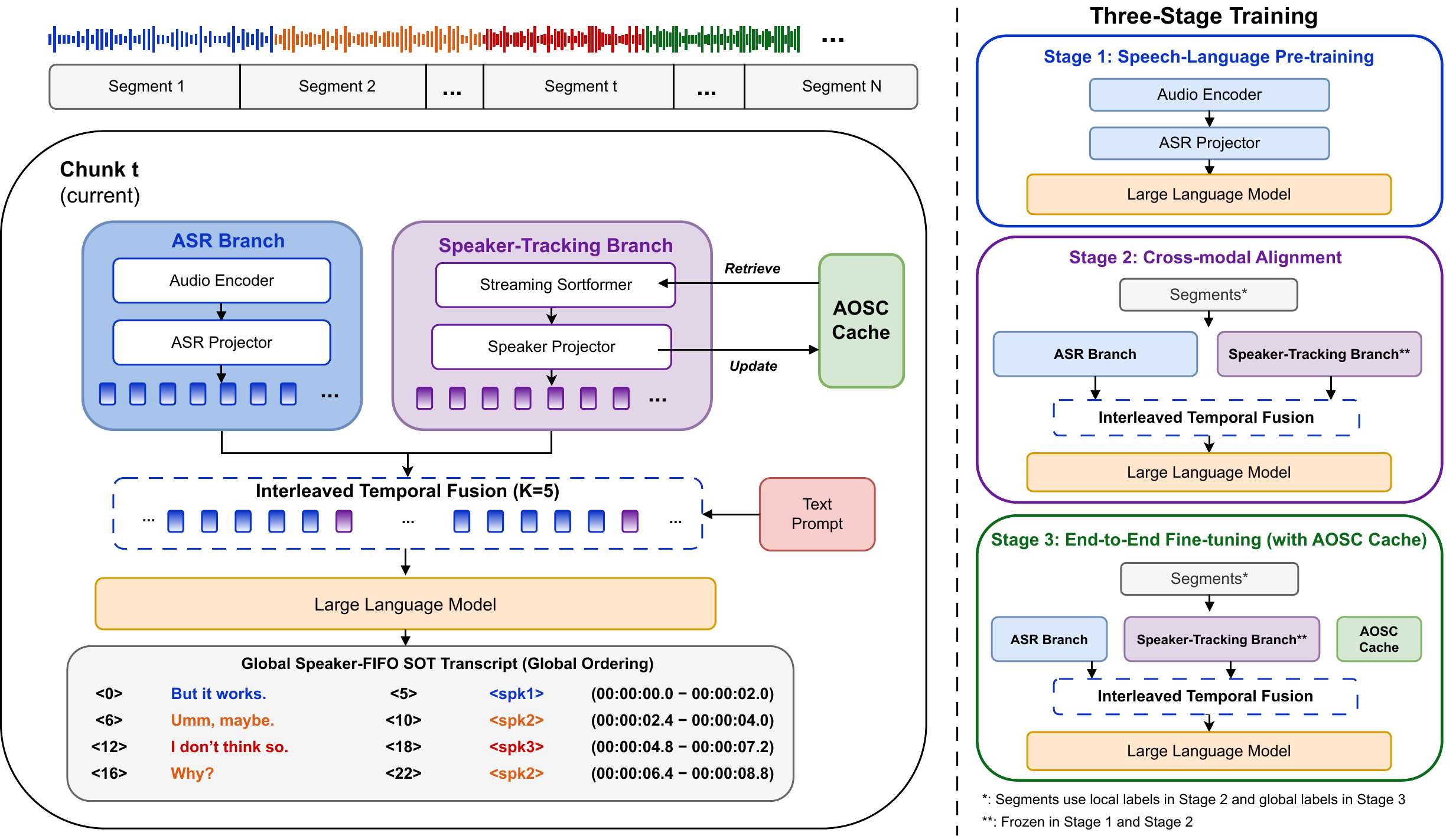}
  \caption{Overview of G-STAR Structure. The left panel illustrates the proposed G-STAR model architecture, which features a dual-branch encoder design comprising ASR and speaker diarization modules to support chunk-wise inference. The right panel outlines the three-stage training strategy for G-STAR, progressing sequentially from Speech-Language Pre-training to local SA-ASR (Cross-modal Alignment) and finally to global SA-ASR (End-to-End Fine-tuning).}
  \label{fig:structure_overview}
\end{figure*}

Overall, these lines of work suggest complementary strengths: Sortformer-style models provide stable tracking cues, whereas SOT-style generation provides a flexible text interface for attributed transcripts. G-STAR (Fig.~\ref{fig:structure_overview}) combines these two perspectives by treating speaker tracking as structured conditioning for generation, which is the bridge from the background reviewed here to the method introduced next.

\endgroup

\section{Method}
\label{sec:method}

\subsection{Problem Setup and Notation}
Let a long-form recording be a waveform $x$ partitioned into $T$ consecutive chunks
$\{x^{(t)}\}_{t=1}^{T}$.
In practice, chunk durations are \emph{approximately stable but not fixed} due to chunking constraints,
utterance boundaries, or buffering policies; thus the number of encoder frames $L_t$ and tracker frames $M_t$
may vary with $t$.

Our goal is \emph{timestamped speaker-attributed ASR with meeting-level consistent speaker identities}.
For each chunk $t$, the model emits a serialized sequence in an SOT-style format:
\begin{equation}
\label{eq:sot_format}
\resizebox{0.85\columnwidth}{!}{
$\mathbf{z}^{(t)} = \big[\tok{t\_st},\ \mathbf{w},\ \tok{t\_ed},\ \tok{spk{=}k}\big]^{\ast},$}
\end{equation}
where $\mathbf{w}$ denotes lexical tokens, and $\tok{spk{=}k}$ denotes a \emph{global} speaker-ID token
that is consistent across all chunks within the same meeting.

G-STAR couples three components in a chunk-wise fashion:
(i) an ASR acoustic branch that produces LLM-conditioned acoustic embeddings,
(ii) an SD/speaker-tracking branch that produces time-synchronous speaker cues, and
(iii) a persistent \emph{Arrival-Order Speaker Cache} (AOSC) that maintains meeting-level speaker identities.
At chunk $t$, the SD branch consumes the previous cache state $\mathcal{C}^{(t-1)}$,
produces speaker cues, and updates the cache to $\mathcal{C}^{(t)}$.
The LLM then performs SOT decoding conditioned on the fused acoustic/speaker embeddings.

\subsection{Chunk-Wise Dual-Branch Modeling with AOSC}
\label{sec:method_chunkwise}

\subsubsection{ASR Acoustic Branch}
Given chunk $x^{(t)}$, an audio encoder $f_{\mathrm{enc}}(\cdot)$ (e.g., Conformer/Whisper-style) produces
frame-level representations:
\begin{equation}
\mathbf{H}^{(t)} = f_{\mathrm{enc}}(x^{(t)}) \in \mathbb{R}^{L_t \times d_h},
\end{equation}
where $L_t$ varies with chunk duration.
A projector $g_1(\cdot)$ maps them into the LLM embedding space:
\begin{equation}
\mathbf{U}^{(t)} = g_1(\mathbf{H}^{(t)}) \in \mathbb{R}^{L_t \times d_{\mathrm{llm}}}.
\end{equation}

\subsubsection{SD Branch and Cache-Conditioned Speaker Cues}
The SD/speaker-tracking module maintains a persistent cache $\mathcal{C}^{(t)}$ across chunks.
For chunk $t$, the tracker takes the current audio and the previous cache state as input,
and outputs frame-synchronous cache-conditioned speaker cues:
\begin{equation}
\label{eq:trk_out}
\begin{aligned}
\mathbf{S}^{(t)}, \mathcal{C}^{(t)} &= f_{\mathrm{trk}}(x^{(t)}, \mathcal{C}^{(t-1)}), \\
\mathbf{S}^{(t)} &\in \mathbb{R}^{M_t \times d_s},
\end{aligned}
\end{equation}
where $M_t$ is the number of tracker frames, also varying with chunk duration and the tracker frame rate.
A projector $g_2(\cdot)$ maps the speaker cues into the LLM embedding space:
\begin{equation}
\mathbf{V}^{(t)} = g_2(\mathbf{S}^{(t)}) \in \mathbb{R}^{M_t \times d_{\mathrm{llm}}}.
\end{equation}

AOSC stores compact speaker evidence ordered by \emph{first-arrival time}, thereby defining a stable
meeting-level indexing rule. When a new speaker appears, AOSC assigns the next arrival-order index;
when a previously seen speaker reappears, AOSC retrieves the corresponding index.
This renders speaker identity \emph{stateful} across chunks and eliminates cross-chunk permutation drift. The detailed cache retrieval, allocation, and update procedure of AOSC is illustrated in Fig.~\ref{fig:streaming_sortformer}, where returning speakers are linked to existing arrival-order slots and newly detected speakers are assigned to the next available slot.

\subsection{Interleaved Temporal Fusion and Global SOT Decoding}
\label{sec:method_fusion_decode}

\subsubsection{Interleaved Temporal Fusion} 
We fuse acoustic embeddings $\mathbf{U}^{(t)}$ and speaker embeddings $\mathbf{V}^{(t)}$ \emph{along time}
by inserting sparse speaker tokens into the acoustic token stream.
Let $K$ be the insertion stride. We construct a fused embedding sequence:
\begin{equation}  
\label{eq:interleave}
\resizebox{0.85\columnwidth}{!}{
$\mathbf{E}^{(t)} = \mathrm{Interleave}(\mathbf{U}^{(t)}, \mathbf{V}^{(t)}; K)
\in \mathbb{R}^{N_t \times d_{\mathrm{llm}}},$}
\end{equation}
where $N_t \approx L_t + \lceil L_t/K \rceil$.
Operationally, for $i=1,\dots,L_t$, we append $\mathbf{U}^{(t)}_i$ and additionally insert a speaker embedding whenever
$i \equiv 0 \ (\mathrm{mod}\ K)$.
When $M_t$ does not match the insertion schedule, we apply a deterministic resampling (nearest-neighbor or linear)
to align speaker frames to insertion locations.
This yields a single time-ordered embedding stream that periodically carries explicit speaker evidence.

\subsubsection{Cache-Consistent SOT Decoding with Global Speaker IDs}
We feed $\mathbf{E}^{(t)}$ to an autoregressive LLM decoder.
Optionally, a text prompt $\mathbf{p}$ is prepended.
The decoder generates the SOT sequence conditioned on the fused embeddings:
\begin{equation}
\label{eq:ar_llm}
\resizebox{0.85\columnwidth}{!}{
$p(\mathbf{z}^{(t)} \mid x^{(t)}, \mathcal{C}^{(t-1)})
= \prod_{m=1}^{|\mathbf{z}^{(t)}|}
p(z^{(t)}_m \mid \mathbf{p}, \mathbf{E}^{(t)}, z^{(t)}_{<m}).$}
\end{equation}
The speaker token $\tok{spk{=}k}$ in Eq.~\eqref{eq:sot_format} corresponds to the $k$-th arrival-order index maintained by AOSC.
The tracker update enables end-to-end meeting-level identity consistency
during chunk-wise inference (Alg.~\ref{alg:gstar}).

\begin{algorithm}[t!]
\caption{G-STAR chunk-wise inference with cache-conditioned tracking and global SOT decoding}
\label{alg:gstar}
\small
\begin{algorithmic}[1]
\REQUIRE Meeting chunks $\{x^{(t)}\}_{t=1}^{T}$; insertion stride $K$; optional prompt $\mathbf{p}$.
\ENSURE Timestamped transcript with meeting-level speaker IDs.
\STATE Initialize AOSC cache $\mathcal{C}^{(0)} \leftarrow \varnothing$ and transcript list $\mathcal{Z}$.
\FOR{$t = 1$ to $T$}
  \STATE Encode acoustic frames: $\mathbf{H}^{(t)} \leftarrow f_{\mathrm{enc}}(x^{(t)})$.
  \STATE Project ASR features: $\mathbf{U}^{(t)} \leftarrow g_1(\mathbf{H}^{(t)})$.
  \STATE Track speakers with history: $(\mathbf{S}^{(t)}, \mathcal{C}^{(t)}) \leftarrow f_{\mathrm{trk}}(x^{(t)}, \mathcal{C}^{(t-1)})$.
  \STATE Maintain arrival-order speaker indices for $\tok{spk{=}k}$.
  \STATE Project speaker cues: $\mathbf{V}^{(t)} \leftarrow g_2(\mathbf{S}^{(t)})$.
  \STATE Resample $\mathbf{V}^{(t)}$ to the ASR insertion grid.
  \STATE Build $\mathbf{E}^{(t)} \leftarrow \mathrm{Interleave}(\mathbf{U}^{(t)}, \mathbf{V}^{(t)}; K)$ in temporal order.
  \STATE Decode $\mathbf{z}^{(t)} \leftarrow \mathrm{LLMDecode}(\mathbf{p}, \mathbf{E}^{(t)})$.
  \STATE Append $\mathbf{z}^{(t)}$ to $\mathcal{Z}$; retain $\mathcal{C}^{(t)}$.
\ENDFOR
\STATE \textbf{return} $\mathrm{Concat}(\mathcal{Z})$.
\end{algorithmic}
\end{algorithm}


\section{Experimental Setups}
\label{sec: experiment}
We design the experiments to answer two complementary questions. The local setting isolates speaker-attributed recognition under controlled utterance-level segmentation, whereas the global setting evaluates full-meeting behavior where segmentation, speaker tracking, and attribution interact over long contexts. This separation is important because local and global results reflect different sources of difficulty.
We implemented our model and conducted experiments using the WEST~\citep{zhang2025westllmbasedspeech} framework.

\subsection{Data}
We trained and evaluated our models on four conversational and meeting-style speech datasets: MLC~\citep{mu2025summary}, AMI~\citep{carletta2005ami}, Fisher~\citep{cieri2004fisher}, and Candor~\citep{reece2023candor}. For MLC, only the English subset was used; for Candor, only the audio modality was adopted, excluding video information.
All training audio was segmented into clips of up to 20 seconds.
For evaluation, two settings were considered: in the local setting, the test set was processed in the same way as the training data; in the global setting, the test set comprised full-length audio, with the model performing chunk-by-chunk inference.

In addition to the public benchmarks, we also include internal data for model training and evaluation. For training, we added an internal Chinese conversational corpus, which is dominated by two-speaker conversations.
The only training-data difference between the internal model and the open-source model is whether internal Chinese conversational corpus and the open-source AISHELL-4~\cite{aishell4} and AliMeeting~\cite{alimeeting} datasets are used. The internal test set is collected from real out-of-domain meeting audio and contains two-speaker, three-speaker, and four-speaker recordings of 0.50, 0.49, and 1.86 hours, respectively.

\subsection{Modeling}
We have detailed the proposed architecture in Section~\ref{sec:method_chunkwise}.
The audio encoder and projector are initialized from FireRed-LLM~\citep{xu2025fireredasr}.
The LLM uses Qwen2-7B-Instruct~\citep{yang2025qwen3} with LoRA weights inherited from FireRed-LLM.
For the SD branch, the speaker diarization module is initialized with Streaming Sortformer\footnote{\href{https://huggingface.co/nvidia/diar_streaming_sortformer_4spk-v2}{\texttt{nvidia/diar\_streaming\_sortformer\_4spk-v2}}}, whereas its projector consists of a 1D convolutional downsampling module with a stride of 5, followed by a two-layer MLP, and is randomly initialized.

\subsection{Training Procedure}
The overall optimization followed three consecutive stages.
\begin{itemize}
  \item First, we performed meeting-style ASR pre-training to adapt the speech encoder--projector--LLM pipeline to conversational and meeting-style audio.
  \item Second, we conducted local SA-ASR training on segmented utterances, where the model learned to generate timestamped transcriptions with speaker-attribution tokens.
  \item Third, we performed global SA-ASR training to adapt the system to meeting-level inference, where speaker identities must remain consistent across chunks rather than being assigned independently within each segment.
\end{itemize}
During this global stage, we also tuned the Sortformer module on 90-second chunks from the same four datasets to improve long-form speaker tracking under realistic meeting-level inference.
This component can be tuned modularly because the diarization module is coupled to the downstream Speech-LLM generator through the SD projector.

\subsection{Training Hyperparameters}
We used packed training examples with a maximum pack size of 12,000 tokens in all Speech-LLM training stages.
The trainable modules were kept consistent across these stages: the ASR projector, the SD projector, and the LoRA adapters.
The LoRA rank, scaling factor, and dropout were fixed to 64, 16, and 0.05, respectively.
We optimized the Speech-LLM components with AdamW using a weight decay of 0.01, $\beta_1=0.9$, $\beta_2=0.95$, and gradient clipping.
For meeting-style ASR pre-training and local speaker-attributed ASR (SA-ASR) training, the learning rate was linearly warmed up with a warmup ratio of 0.01 to a peak value of $5\times10^{-5}$ and then decayed with cosine annealing; each of the two stages was trained for 20,000 steps.
For global (meeting-level) SA-ASR training, we kept the same pack size and learning-rate schedule, but reduced the peak learning rate to $2\times10^{-5}$ and trained for 5,000 steps.
Timestamps and speaker labels were added as special LLM tokens.
We optimized the generation objective with hierarchical cross-entropy, assigning a $1.5\times$ loss weight to timestamp tokens and a $2\times$ loss weight to speaker-label tokens.
For Sortformer tuning in the global stage, we used AdamW with a learning rate of $1\times10^{-4}$ and a batch size of 4, and trained for 5 epochs on 90-second segments constructed from AMI, Fisher, MLC, and Candor.

\begin{table*}[!t]
  \centering
  \caption{Performance comparison in the local setting (audio up to 20s) with oracle VAD/segmentation, reported in cpWER/DER (\%). The NVIDIA default cascade is shown as a two-row group: Sortformer for SD/DER and Parakeet for ASR/cpWER. DER is computed with collar 0. \textbf{Bold} indicates the best result.}
  \label{tab:local_cpwer_der}
  \setlength{\tabcolsep}{4pt}
  \renewcommand{\arraystretch}{1.08}
  \resizebox{\textwidth}{!}{%
  \begin{tabular}{llcccc}
    \toprule
    \multirow{2}{*}{\textbf{Group}} &
    \multirow{2}{*}{\textbf{Component}} &
    \multicolumn{1}{c}{\textbf{AMI}} &
    \multicolumn{1}{c}{\textbf{Fisher}} &
    \multicolumn{1}{c}{\textbf{MLC}} &
    \multicolumn{1}{c}{\textbf{Candor}} \\
    \cmidrule(lr){3-3} \cmidrule(lr){4-4} \cmidrule(lr){5-5} \cmidrule(lr){6-6}
    & & \textbf{cpWER/DER$\downarrow$}
    & \textbf{cpWER/DER$\downarrow$}
    & \textbf{cpWER/DER$\downarrow$}
    & \textbf{cpWER/DER$\downarrow$} \\
    \midrule
    \multirow{2}{*}{NVIDIA cascade} & Sortformer (SD)~\citep{park2024sortformer} & -- / 29.87 & -- / 18.33 & -- / 17.76 & -- / 30.92 \\
    & Parakeet (ASR)~\citep{wang25y_interspeech} & \textbf{24.62} / -- & 27.73 / -- & 25.90 / -- & 27.44 / -- \\
    \midrule
    \multirow{2}{*}{E2E SA-ASR} & Vibevoice-ASR~\citep{peng2026vibevoice} & 30.51 / 31.99 & 15.18 / 17.68 & 21.74 / 14.01 & 22.12 / 30.89 \\
    & MOSS-Diarizen~\citep{ai2026mosstranscribediarizetechnical} & 25.13 / 32.20  & 11.69 / 21.61 & 14.16 / 10.58 & 16.38 / 31.76 \\
    \midrule
    \textsc{G-STAR} & Ours & 24.86 / \textbf{19.00} & \textbf{10.29} / \textbf{8.18} & \textbf{13.90} /  \textbf{6.49} & \textbf{14.54} / \textbf{17.56} \\
    \bottomrule
  \end{tabular}%
  }
\end{table*}

\subsection{Evaluation Setup}
Unless otherwise noted, all reported inference uses chunk-wise decoding.
In the local setting, all systems are evaluated on utterance-level inputs of up to 20 seconds with oracle VAD/segmentation, which isolates the back-end speaker-attributed recognition capability from segmentation errors. In the global setting, systems are evaluated on full-meeting recordings with each model's own VAD or segmentation front-end, which better reflects realistic long-form usage but also means that VAD behavior is part of the evaluated system.
Unless otherwise specified, the collar for DER is set to 0.
When hallucinated hypothesis segments appear, we score them consistently for all systems: the hallucinated outputs are treated as errors and are reflected in cpWER and DER.

To assess whether G-STAR offers benefits beyond post-hoc integration, we implement a cascaded pipeline as the closest non-end-to-end counterpart. For this baseline, we used the same Speech-LLM back-end, FireRedASR-LLM, for the ASR component. A VAD model\footnote{\url{https://github.com/snakers4/silero-vad}} was employed to remove silence segments and generate input utterances for ASR while preserving global timestamps. As FireRedASR-LLM does not provide word-level timestamps, we applied a pre-trained CTC-based forced alignment model\footnote{\href{https://huggingface.co/MahmoudAshraf/mms-300m-1130-forced-aligner}{\texttt{MahmoudAshraf/mms-300m-1130-forced-aligner}}} to align the ASR outputs and obtain word-level timestamps. For speaker diarization, we adopted the same Sortformer-style front-end as \textbf{G-STAR} to handle long-form speech and produce global speaker timestamps. Finally, the global word-level timestamps from ASR were integrated with the speaker timestamps to generate the final speaker-attributed transcription with global timing information.

For the internal test set, we also compare with a cluster-based internal baseline, which performs segment-level recognition followed by global speaker clustering to produce speaker-attributed transcripts.
\section{Evaluation}
\label{sec: eval}

Throughout this paper, we use \emph{meeting-level} and \emph{global} to refer to speaker identity consistency over the full recording.

\subsection{Chunk-Level Evaluation}

The chunk-level evaluation in Table~\ref{tab:local_cpwer_der} uses utterance-level inputs of up to 20 seconds with oracle VAD/segmentation, providing a controlled test of speaker-attributed recognition without front-end segmentation variability.
This setting isolates the backend capability of each system, namely whether it can jointly model lexical content, speaker attribution, and local temporal structure when reliable input boundaries are available.

The Sortformer and Parakeet rows should be read together as the strong cascade: Sortformer reports the SD/DER component, while Parakeet reports the ASR/cpWER component.
Under this local protocol, G-STAR consistently reduces DER relative to Sortformer, showing stronger speaker-attribution behavior within short chunks.
At the same time, it remains comparable to Parakeet on AMI and achieves lower cpWER on Fisher, MLC, and Candor, suggesting that the injected speaker cues do not compromise lexical recognition. Compared with strongend-to-end Speech-LLM baselines such as VibeVoice-ASR~\citep{peng2026vibevoice} and MOSS-Diarizen~\citep{ai2026mosstranscribediarizetechnical}, G-STAR improves both metrics on all datasets.
These results indicate that explicit cache-conditioned speaker cue fusion strengthens the LLM's ability to generate speaker-attributed transcripts, rather than relying only on the transcription backbone or on post-hoc diarization-style attribution.

\subsection{Meeting-Level Evaluation}

\begin{table*}[!t]
  \centering
  \caption{Global (meeting-level) performance comparison, reported in cpWER/DER (\%). The NVIDIA default cascade is shown as separate SD and ASR components; the controlled cascade uses the same main components as G-STAR but applies late fusion. DER is computed with collar 0. \textbf{Bold} indicates the best result.}
  \label{tab:global_fisher_mlc_cpwer_der}
  \setlength{\tabcolsep}{5pt}
  \small
  \begin{resizebox}{\textwidth}{!}{
  \begin{tabular}{llcccc}
    \toprule
    \multirow{2}{*}{\textbf{Group}} &
    \multirow{2}{*}{\textbf{Component}} &
    \textbf{Fisher} &
    \textbf{MLC} &
    \textbf{Candor} &
    \textbf{AMI} \\
    \cmidrule(lr){3-3} \cmidrule(lr){4-4} \cmidrule(lr){5-5} \cmidrule(lr){6-6}
    & & \textbf{cpWER/DER$\downarrow$} & \textbf{cpWER/DER$\downarrow$} & \textbf{cpWER/DER$\downarrow$} & \textbf{cpWER/DER$\downarrow$} \\
    \midrule
    \multirow{2}{*}{NVIDIA cascade} & Sortformer (SD)~\citep{park2024sortformer} & -- / \textbf{15.21} & -- / 21.92 & -- / 18.03 & -- / \textbf{28.35} \\
    & Parakeet (ASR)~\citep{wang25y_interspeech} & 24.41 / -- & 31.03 / -- & 26.92 / -- & 35.70 / -- \\
    \midrule
    Speech-LLM & Vibevoice-ASR~\citep{peng2026vibevoice} & 25.03 / 27.15 & 25.41 / 19.83 & 27.24 / 25.68 & 34.19 / 39.95 \\
    Controlled & Late-fusion cascade & 21.01 / 23.41  & 23.18 / 21.38 & 17.62 / \textbf{17.67} & 39.52 / 37.63\\
    \textsc{G-STAR} & Ours & \textbf{16.44} / 16.85  & \textbf{17.15} / \textbf{14.25} & \textbf{15.17} / 24.89 & \textbf{30.85} / 32.23\\
    \bottomrule
  \end{tabular}
  }\end{resizebox}
\end{table*}

As shown in Table~\ref{tab:global_fisher_mlc_cpwer_der}, \textsc{G-STAR} achieves the best meeting-level cpWER on Fisher, MLC, Candor, and AMI.
The NVIDIA cascade rows report its SD and ASR components separately, while our controlled cascade uses the same main tracker/backbone design as G-STAR but performs late fusion between ASR and speaker timestamps.
Compared with this controlled cascade, G-STAR obtains consistently lower cpWER, showing that the gain comes from tracking-conditioned generation rather than only from stronger components.
The DER results reveal the expected trade-off: dedicated diarization systems can still be stronger for diarization purity, whereas G-STAR prioritizes speaker-attributed transcription with cache-consistent global labels.

We also conduct an extended evaluation on an internal out-of-domain test set to verify that the observed trend generalizes to practical meeting scenarios, including two-speaker and three-to-four-speaker configurations. Table~\ref{tab:scene_cpwer_der} shows that our system, after further fine-tuning on a small amount of internal conversational data, consistently outperforms both the internal baseline and VibeVoice-ASR in terms of DER and cpWER.

\begin{table}[!t]
  \centering
    \caption{Global SA-ASR performance on the internal test set, reported in cpWER/DER (\%). DER is computed with collar 0.5.}
  \label{tab:scene_cpwer_der}
  \setlength{\tabcolsep}{5pt}
  \renewcommand{\arraystretch}{1.08}
  \small
  \resizebox{\columnwidth}{!}{%
  \begin{tabular}{lccc}
    \toprule
    \multirow{2}{*}{\textbf{System}} &
    \textbf{2 speakers} &
    \textbf{3--4 speakers} &
    \textbf{Avg.} \\
    \cmidrule(lr){2-2} \cmidrule(lr){3-3} \cmidrule(lr){4-4}
    & \textbf{cpWER/DER$\downarrow$} & \textbf{cpWER/DER$\downarrow$} & \textbf{cpWER/DER$\downarrow$} \\
    \midrule
    Vibevoice-ASR~\citep{peng2026vibevoice} & 11.10 / 14.83 & 54.48 / 38.33 & 47.64 / 34.62 \\
    Cluster-based Pipeline& 23.56 / 14.20 & 41.14 / 30.76 & 38.37 / 28.15 \\
    G-STAR & \textbf{10.42} / \textbf{4.86} & \textbf{38.85} / \textbf{28.59} & \textbf{34.37} / \textbf{24.88} \\
    \bottomrule
  \end{tabular}%
  }
\end{table}

\newcommand{\cmark}{\checkmark}
\newcommand{\xmark}{$\times$}

\subsection{Ablation Study}

After establishing the main local and global results, we analyze why the proposed architecture helps. The ablation focuses on two modeling choices that directly control how speaker information enters the LLM: interleaved temporal fusion and hierarchical cross-entropy weighting.

Table~\ref{tab:gstar_ablation_local_singlecol} shows that the two components are complementary rather than redundant.
Interleave fusion reduces both cpWER and DER, with a larger gain on cpWER, suggesting that periodically injecting speaker cues helps the LLM generate the correct lexical sequence together with structure-critical tokens such as speaker and timestamp markers.
In contrast, hierarchical CE yields a clearer improvement on DER while leaving cpWER largely unchanged, indicating that reweighting timestamp and speaker-label tokens mainly sharpens temporal boundaries and speaker turn segmentation.
The full model combines these effects, improving token-level attribution through interleave fusion and diarization-sensitive structure prediction through hierarchical CE.

\begin{figure}[!t]
  \centering
  \includegraphics[width=\columnwidth]{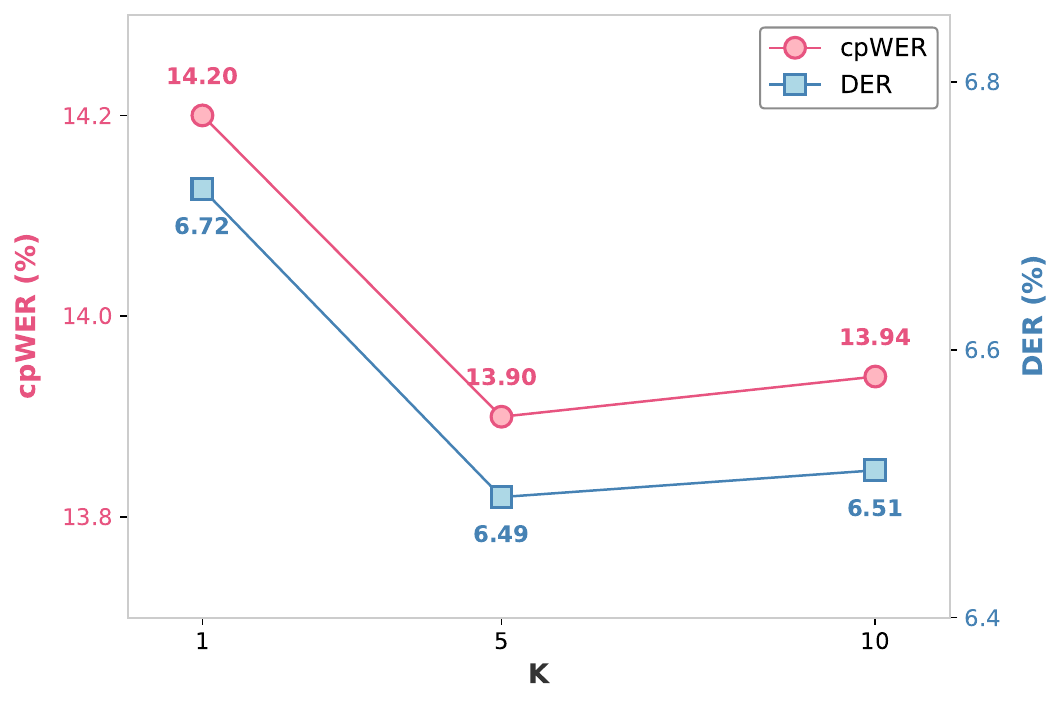}
  \caption{Effect of the interleave ratio $K$ on MLC SA-ASR performance in the local setting. Smaller $K$ injects speaker cues more frequently, while larger $K$ provides sparser speaker conditioning.}
  \label{fig:interleave_ratio_analysis}
\end{figure}

We further analyze the effect of the interleave ratio, which controls the relative frequency with which speaker cues are injected into the LLM token stream during fusion.
As shown in Figure~\ref{fig:interleave_ratio_analysis}, an overly dense injection can interfere with lexical modeling by increasing the proportion of non-lexical conditioning tokens, while an overly sparse injection may provide insufficient speaker-tracking guidance.
The results suggest that a moderate interleave ratio achieves the best balance between speaker attribution and transcription accuracy, confirming that the benefit of interleave fusion depends not only on whether speaker cues are used, but also on how frequently they are integrated.

\begin{table}[!b]
  \centering
  \caption{Ablation study of model configuration under local setting.} 
  \label{tab:gstar_ablation_local_singlecol}
  \setlength{\tabcolsep}{4pt}
  \renewcommand{\arraystretch}{1.02}
  \scriptsize
  \resizebox{\columnwidth}{!}{%
  \begin{tabular}{cc c c c}
    \toprule
    \multirow{2}{*}{\makecell[c]{\textbf{Hierarchical}\\\textbf{CE loss}}} &
    \multirow{2}{*}{\makecell[c]{\textbf{Interleave}\\\textbf{Temporal Fusion}}} &
    \textbf{AMI} & \textbf{Fisher} & \textbf{Candor} \\
    \cmidrule(lr){3-3}\cmidrule(lr){4-4}\cmidrule(lr){5-5}
    & &
    \makecell[c]{\textbf{cpWER/DER}$\downarrow$} &
    \makecell[c]{\textbf{cpWER/DER}$\downarrow$} &
    \makecell[c]{\textbf{cpWER/DER}$\downarrow$} \\
    \midrule
    $\times$ & $\checkmark$ & 26.33 / 21.06 & 10.88 / 10.24 & 14.97 / 20.21 \\
    $\checkmark$ & $\times$ & 28.63 / 21.28 & 14.23 / 9.02  & 18.30 / 18.10 \\
    \midrule
    $\checkmark$ & $\checkmark$ &
    \textbf{24.86} / \textbf{19.00} &
    \textbf{10.29} / \textbf{8.18} &
    \textbf{14.54} / \textbf{17.56} \\
    \bottomrule
  \end{tabular}%
  }
\end{table}


\section{Conclusion}
We presented \textsc{G-STAR}, an end-to-end framework for timestamped speaker-attributed ASR under chunk-wise long-form inference. G-STAR uses cache-conditioned Sortformer-style tracking to provide slot-aligned speaker cues to the LLM, enabling lexical tokens, timestamps, and global speaker IDs to be generated under a shared conditioning context. Experiments across local and meeting-level settings show improved speaker-attributed transcription, and ablations confirm the roles of interleaved cue fusion and hierarchical CE. These results suggest that tracking-conditioned generation is a practical alternative to late-fusion cascades.

Beyond the empirical gains, the main implication of G-STAR is that speaker attribution can be treated as a conditioning signal during generation rather than as a post-processing decision made after ASR. The cache preserves a compact notion of who has appeared, the interleaved fusion mechanism exposes this information to the language model at temporally meaningful points, and the global SOT interface converts the resulting context into a single readable transcript. This combination is especially useful for long-form conversations. Overall, the framework offers a unified path for jointly improving transcription, timestamping, and speaker consistency while remaining compatible with modular training and chunk-wise deployment.

\newpage

\section*{Limitations}

Although \textsc{G-STAR} is designed around a cache-conditioned speaker tracker and is compatible with chunk-wise inference, this work does not fully evaluate a strictly streaming deployment. Future work will extend the system to real-time streaming settings and analyze latency, memory usage, and cache update stability under online constraints. Another limitation is data scale: the current experiments cover several public and internal meeting-style datasets, but larger and more diverse training data may further improve robustness across acoustic conditions, speaking styles, languages, and speaker configurations. Scaling both the Speech-LLM training data and the speaker-tracking supervision is therefore an important direction for improving generalization.


\bibliography{mybib}






\end{document}